\documentstyle[11pt,moriond,epsfig]{article}

\def\mco{\multicolumn}

\def\be{\begin{equation}}
\def\ee{\end{equation}}
\def\bea{\begin{eqnarray}}
\def\eea{\end{eqnarray}}
\def\M{{\cal M}}

\newcommand{\gras}[1]{\mbox{\boldmath $#1$}}
\begin{document}
\vspace*{4cm}
\title{DETECTION OF GRAVITATIONAL WAVES FROM INSPIRALING COMPACT BINARIES
USING NON-RESTRICTED POST-NEWTONIAN APPROXIMATIONS}

\author{ALICIA M. SINTES and  ALBERTO VECCHIO}

\address{
Max-Planck-Institut f\"ur Gravitationsphysik
(Albert-Einstein-Institut) \\
Schlaatzweg 1, 14473 Potsdam, Germany}

\maketitle\abstracts{
The set up of matched filters for the detection of gravitational waves from 
in-spiraling compact binaries is usually carried out using
the restricted post-Newtonian approximation: the 
filter phase is modelled including post-Newtonian corrections,
whereas the amplitude is retained at the  Newtonian order.
Here we investigate the effects of the introduction of post-Newtonian 
corrections
also to the amplitude and we discuss some of the implications for 
signal detection and parameter estimation.}

\section{Introduction}

\index It has long been recognized that  binary systems of compact objects are
important sources of gravitational waves (GW) both for the
ground based interferometric detectors 
currently under construction \cite{Thorne97,Cu93} and 
future space-based interferometers like LISA \cite{lisa}.

In this paper, we consider the in-spiral phase of the coalescence of
binary systems in circular orbit using post-Newtonian (PN)
approximations to general relativity \cite{Bl95}. 

The detection of in-spiral signals is carried out by cross-correlating
the detector output with a discrete set of filters \cite{Sa91,Ow96}, usually
computed within the so-called restricted 
post-Newtonian approximation \cite{Sa91}:
PN corrections are taken into account in the phase of the wave-form,
whereas the amplitude is retained at the lowest Newtonian order.
Thus, one discards all multipole components except the quadrupole one.
Such simplification of the filter structure is believed to have 
negligible effects on the detection performances and is
not expected to affect appreciably the  statistical 
errors in the estimation of the source parameters.

Here we investigate the effects of the introduction of PN corrections
also to the amplitude of the wave-form 
 and discuss some implications for 
signal detection and parameter estimation. For sake of simplicity 
we will assume negligible spins and, as it is always the case for ground based
experiments, circular orbits.

\section{The wave-form}

The signal produced at the output of an interferometric detector
by a gravitational wave of polarization amplitudes $h_+$ and $h_{\times}$ can 
be written as
\be
h(t)=F_+ h_+(t) + F_{\times}h_{\times}(t) \ ,
\label{h1}
\ee
where $F_+$ and $F_{\times}$ are the so-called beam pattern functions
of the detector \cite{Th300}; they depend on the 
location of the source in the sky $(\theta,\phi)$ and the polarization angle $\psi$.
If we consider the in-spiral of a binary system of masses $m_1$ and $m_2$,
$h_+$ and $h_{\times}$ read \cite{Bl96}
\be
h_{+,\times}= {2m\eta\over r}x\left\{H_{+,\times}^{(0)}
+x^{1/2}H_{+,\times}^{(1/2)}+x H_{+,\times}^{(1)} +x^{3/2}H_{+,\times}^{(3/2)}
+x^2 H_{+,\times}^{(2)}+ \cdots \right\} \ ,
\label{m4}
\ee
where $x\equiv (m\omega)^{2/3}$,
$\omega$ is the system orbital frequency and $r$ the source distance;
$m=m_1+m_2$, $\mu=m_1m_2/m$, 
$\eta=\mu/m$ and ${\cal M}=\mu^{3/5}m^{2/5}=m\eta^{3/5}$
are the total mass, the reduced mass, the symmetric mass ratio and 
the chirp mass, respectively. 

The lower terms of the PN expansion for the plus and 
cross polarization are given by
\cite{Bl96}
\bea
H_+^{(0)}&=& -(1+c^2)\cos \Phi \ , \\
H_+^{(1/2)}&=& -s\, {\sqrt{1-4\eta}\over 8}
\left[ (5+c^2)\cos({1 \over 2}\Phi)-9(1+c^2)\cos({3 \over 2}\Phi)\right] \ , \\
H_{\times}^{(0)}&=& -2c \, \sin \Phi \ , \\
H_{\times}^{(1/2)}&=& -{3\over 4}\, s \, c\, \sqrt{1-4\eta}\left[
\sin({1 \over 2}\Phi)- 3 \sin ({3 \over 2}\Phi)\right] \ ,
\eea
where $c=\cos \iota$ and $s=\sin \iota$; $\iota$ is the angle between 
the direction of the source and the orbital angular momentum,
 and $\Phi$ is twice the orbital phase.
 
 If one  considers only the first term in (\ref{m4}),  corresponding
 to the Newtonian one, one gets the restricted PN approximation.
 Our goal is to study  the full 2 PN wave-form; here we
 will  present some preliminary results, where
 the  0.5 PN corrections to the amplitude are taken into account,
 keeping however the phase at 2 PN order. 
Although this is a simplified and, to some
 extent, arbitrary choice of signal, all new features of the 
 wave-form are introduced, in particular more 
 information about the masses and the
 position of the source.
 
 Considering the amplitude through 0.5-PN order, the GW output at
 the detector can be written as
\be
h(t)={2m\eta\over r}(m\pi F)^{2/3}\left\{h^{(0)}(t)
+(m\pi F)^{1/3}h^{(1/2)}(t)\right\} \ ,
\ee
where $F$ is the quadrupole gravitational wave frequency,
i.e., $d\Phi/dt=2\pi F$, and
\be
h^{(0)}(t)=-\sqrt{F_+^2(1+c^2)^2+F_{\times}^2 4c^2}\, \cos (\Phi +\varphi_{(0)})
 \ ,
\ee

\be
\varphi_{(0)}=\arctan\left\{ {-2 c F_{\times}\over (1+c^2)F_+} \right\} \ ,
\ee
\bea
h^{(1/2)}(t)=s {\sqrt{1-4\eta}\over 4}& & 
\left[ -\sqrt{F_+^2{(5+c^2)^2\over 4}+F_{\times}^29 c^2 } \,
\cos({1\over 2}\Phi + \varphi_{(1/2)}) \right. \\
& & \left. + {9\over 2}\sqrt{F_+^2(1+c^2)^2+F_{\times}^2 4c^2}\,
 \cos({3\over 2}\Phi + \varphi_{(0)}) \right] \ , \nonumber
\eea

\be
\varphi_{(1/2)}=\arctan\left\{ {-6 c F_{\times}\over (5+c^2)F_+} \right\} \ .
\ee

The Fourier transform of $h(t)$, 
calculated using the stationary phase approximation 
\cite{Sa91,Cu94}, reads:
\be
\tilde h(\nu)=-\sqrt{F_+^2(1+c^2)^2+F_{\times}^24c^2}\sqrt{{5\pi\over 96}}
{(\pi \nu)^{-7/6}\over r}\M^{5/6} \exp \left[i\left( 2\pi\nu t_c +
\Xi(\nu) -\varphi_{(0)}-{\pi\over 4}\right) \right]\, \Lambda
\ee
where
\bea
 \Lambda= &  & 1+{s\over 4}\sqrt{1-4\eta}
 \sqrt{{F_+^2(5+c^2)^2/4+F_{\times}^29c^2\over
  F_+^2(1+c^2)^2+F_{\times}^2 4c^2} }(\pi m \nu)^{1/3}
  \left({1\over 2}\right)^{1/3}\times \nonumber\\
 & & \exp\left[i\left( {1\over 2}\Xi(2\nu)
  -\Xi(\nu) + \varphi_{(0)}- \varphi_{(1/2)} \right) \right]  \nonumber\\
& &-s {9\over 8}\sqrt{1-4\eta}\, (\pi m \nu)^{1/3}\left({3\over 2}\right)^{1/3}
\exp\left[i\left({3\over 2}\Xi\left({2\over 3}\nu\right) -\Xi(\nu) 
\right) \right]  \ ,
\eea
and
\bea
\Xi(\nu)=-\phi_c +{3\over 4}(8\pi\M \nu)^{-5/3}& & \left[ 1+
{20\over 9}\left( {743\over 336}+{11\over 4}\eta\right)(\pi m \nu)^{2/3}
-16\pi(\pi m \nu)\right. \\
& &\left.  + 10\left({3058673\over 1016064} + {5429\over 1008}\eta
+{617 \over 144}\eta^2
\right)(\pi m \nu)^{4/3} \right] \ .\nonumber
\eea

In the restricted PN approximation $\Lambda= 1$. Now it contains
two additional contributions related to the 0.5 PN  corrections to the
amplitude.  Notice the dependency of these two terms on  $\sin \iota$
and $\sqrt{1 - 4\eta}$: the departure from the value 
$\Lambda = 1$ increases as $\iota \rightarrow \pi/2$
and $\eta \rightarrow 0$.

\section{Formalism}

We denote by $h$ the \lq \lq true" GW signal 
and $u({\gras \theta})$ the family of templates,
as a function of the parameter vector 
${\gras \theta}=(t_c,\phi_c,{\gras \lambda})$.
The signal to noise ratio SNR, for optimal filtering, is defined 
as \cite{Cu94}
\be
{\rm SNR}=\sqrt{\left(h\vert h\right)} \ ,
\ee
where $(\, \vert \, )$ denotes the usual inner product.
The fraction of SNR obtained by 
cross-correlating a template $u({\gras \theta})$
with $h$ is given by the ambiguity function
 \be
 {\cal A}({\gras \theta})={\left(h\vert u({\gras \theta})\right)\over 
 \sqrt{\left(h\vert h\right)\, \left(u({\gras \theta})\vert 
 u({\gras \theta})\right)}} \,,
 \ee
which depends on the choice of ${\gras \theta}$.
The maximum of the ambiguity function over the whole parameter
space is defined as the  fitting factor \cite{Ap95}
\be
FF=\renewcommand{\arraystretch}{.6}
\begin{array}[t]{c}{\textstyle\max}\\{\scriptstyle{\gras \theta}}\end{array}
{\left(h\vert u({\gras \theta})\right)\over 
 \sqrt{\left(h\vert h\right)\, \left(u({\gras \theta})\vert 
 u({\gras \theta})\right)}} \ .
\ee 
  The fitting factor  is a measure of how well any chosen family of
templates fits the signal $h$. 
The maximization  of the  ambiguity function
over the extrinsic parameters $\phi_c$ and $t_c$,
phase and time of coalescence, is the so-called match
\be
M({\gras \lambda}_1, {\gras \lambda}_2)=\renewcommand{\arraystretch}{.6}
\begin{array}[t]{c}{\textstyle\max}\\{\scriptstyle\phi_c,t_c}\end{array}
{\left(h({\gras \lambda}_1)\vert u({\gras \lambda}_2)\right)\,
 e^{i(2 \pi  f t_c-\phi_c)}\over 
 \sqrt{\left(h({\gras \lambda}_1)\vert h({\gras \lambda}_1)\right)\, 
 \left(u({\gras \lambda}_2)\vert u({\gras \lambda}_2)\right)}} \ .
\ee
For the set up of the bank of filters, one sets a 
 minimal match \cite{Ow96,Ow98}  as the match 
between signal and template in the case where  the signal is equidistant
from all the nearest templates.

\section{Results}

In the results presented in the following, the signal $h$ is computed according
to Eq. (13)-(15), whereas the template 
wave-forms are calculated within the usual 
restricted PN approximation, 
corresponding to $\Lambda = 1$ in Eq. (14). The noise
curve is the one corresponding to the initial LIGO configuration
\cite{Ow98}. 
\begin{table}[t]
\caption{Comparison of the values of the
fitting factors $FF$, and the  ratios ${\cal R}=\sqrt{(u\vert u)/(h\vert h)}$
for the  pair of masses 0.1-10 $M_\odot$, and 1.4-10 $M_\odot$,
 for different orientations in the sky 
and polarization angles.}
\vspace{0.4cm}
\begin{center}
\begin{tabular}{|l|c|rr|rr|}
\hline
 & & \mco{2}{|c|}{ 0.1-10 $M_\odot$} & \mco{2}{|c|}{ 1.4-10 $M_\odot$}\\
Orientation-polarization & $\iota=\widehat{{\bf L N}}$ &  $FF$ &
${\cal R}$ & $FF$ & ${\cal R}$ \\ \hline

$\forall \, \theta , \, \phi , \,\psi$ & $ \pi/2$ &
 0.9319 & 0.9319 &  0.9549 & 0.9560\\ \hline 
$\theta=\pi/6$, $\phi=\pi/4$, $\psi=\pi/4$ & $ \pi/3$ &
 0.9495 & 0.9495 &  0.9669 & 0.9677\\
$\theta=\pi/6$, $\phi=\pi/4$, $\psi=\pi/6$ & $ \pi/3$ &
 0.9498 & 0.9498&  0.9671 & 0.9681\\
$\theta=\pi/6$, $\phi=\pi/4$, $\psi=0$ & $ \pi/3$ &
  0.9516 & 0.9516  & 0.9683 & 0.9691 \\ \hline 
$\theta=\pi/6$, $\phi=\pi/4$, $\psi=\pi/4$ & $ \pi/4$ &
  0.9662 & 0.9662  &  0.9780 & 0.9787\\
$\theta=0$, $\phi=0$, $\psi=0$ & $ \pi/4$ &
  0.9662 & 0.9662 &  0.9780 & 0.9787\\
$\theta=\pi/6$, $\phi=\pi/4$, $\psi=\pi/6$ & $ \pi/4$ &
   0.9664 & 0.9664 & 0.9782 & 0.9789\\ \hline 
$\theta=\pi/6$, $\phi=\pi/4$, $\psi=\pi/4$ & $ \pi/6$ &
 0.9829 & 0.9829  & 0.9890 & 0.9895\\
$\theta=0$, $\phi=0$, $\psi=0$ & $ \pi/6$ &
 0.9829 & 0.9829 &  0.9890 & 0.9895\\
$\theta=\pi/6$, $\phi=\pi/4$, $\psi=\pi/6$ & $ \pi/6$ &
   0.9829 & 0.9829&  0.9890 & 0.9895\\
$\theta=\pi/6$, $\phi=\pi/4$, $\psi=0$ & $\pi/6$ &
0.9831 & 0.9831 & 0.9891 & 0.9896\\ \hline
\end{tabular}
\end{center}
\label{table4}
\end{table}

\begin{table}[b]
\caption{Fitting factors  and 
the ratios  of SNR
for $\iota=\pi/2$.}
\vspace{0.4cm}
\begin{center}
\begin{tabular}{|c|c|r|r||c|c|r|r|}
\hline
$m_1/M_\odot$ & $m_2/M_\odot$ & $FF$ & ${\cal R}$ 
&$m_1/M_\odot$ & $m_2/M_\odot$ & $FF$ & ${\cal R}$
  \\ \hline
50 & 1.4 &  0.9012 & 0.8864 & 
25 & 1.4 &  0.9139& 0.9184 \\
20 & 5 &    0.9602 & 0.9373&
15 & 1.4 &  0.9360 & 0.9362 \\
10  & 9 &  0.9997 & 0.9995 &
10  & 8 &    0.9987 & 0.9984 \\
10  & 5 &  0.9892 & 0.9880 & 
10  & 1.4 &    0.9549 & 0.9560  \\
10  & 1.0 &   0.9487 & 0.9493 &

10  & 0.75 &  0.9446 & 0.9438 \\
10 & 0.5  &  0.9399 & 0.9398  &
10 & 0.25 &   0.9350 & 0.9352 \\
10 & 0.1 &  0.9319 & 0.9319 &
1.4 & 1.0 &  0.9991 & 0.9991 \\
1.4 & 0.75 &  0.9974 & 0.9974 &
1.4 & 0.5 &    0.9941 & 0.9941 \\
1.4 & 0.25 &  0.9885 & 0.9885 &
1.4 & 0.1  & 0.9834 & 0.9834 
\\ \hline
\end{tabular}
\end{center}
\label{table6}
\end{table}

In table \ref{table4}, we give the fitting factors and the
ratios of SNR, ${\cal R}=\sqrt{(u\vert u)/(h\vert h)}$, for 
the pair of masses 0.1-10 $M_\odot$ and 1.4-10 $M_\odot$, for
different orientation and polarization angles.
We note that the fitting factor reaches a minimum for $\iota= \pi/2$
as expected, and monotonically increases as $\iota \rightarrow 0$
or $\iota \rightarrow \pi$; it depends rather weakly on $\theta$, $\phi$
and $\psi$.

For a fix position in the sky and $\iota=\pi/2$,  we calculate then the
fitting factors for different mass pairs. The fitting factor
 varies from 0.87 to 
1.0; $FF$ gets smaller as $m$ increases and/or $\eta$ decreases,
see table \ref{table6}.

It is now interesting to investigate the loss of SNR
if one uses a restricted PN bank of filters to detect signals that include
PN amplitude terms. The discrete mesh of filters is
normally generated in such a way
that
 any  signal in the restricted PN plane produces a match
$M (u({\gras \lambda}_1), u({\gras \lambda}_2))$ is always larger
than a minimum value, usually set to 0.97.
Assuming that the \lq \lq true" GW signal, 
$h$, is includes PN corrections to the 
amplitude, so that it lies outside the
template space, we want to quantify the match 
$M (h({\gras \lambda}_1), u({\gras \lambda}_2))$ 
between the PN signal and the nearest 
restricted PN template.
 
\begin{table}[h]
\caption{ Fitting factors and match when the GW signal 
and the template have different parameter values. Results for
0.1-10$M_{\odot}$.}
\vspace{0.4cm}
\begin{center}
\begin{tabular}{|lrccc|}
\hline
$\iota$ & $FF$ & $M (u({\gras \lambda}_1), u({\gras \lambda}_2))$
 & $M (h({\gras \lambda}_1), u({\gras \lambda}_2))$ &
$FF\times M (u({\gras \lambda}_1), u({\gras \lambda}_2))$\\ \hline
$\pi/2$ & 0.9319 & 0.9319 & 0.8684 & 0.8684 \\
        &        & 0.9664 & 0.9006 & 0.9006 \\
        &        & 0.9693 & 0.9032 & 0.9033 \\
        &        & 0.9779 & 0.9113 & 0.9113 \\
        &        & 0.9826 & 0.9157 & 0.9157 \\
 \hline      
$\pi/3$ & 0.9495 & 0.9319 & 0.8847 & 0.8848 \\
        &        & 0.9664 & 0.9176 & 0.9176 \\
        &        & 0.9693 & 0.9203 & 0.9203 \\
        &        & 0.9779 & 0.9285 & 0.9285 \\
        &        & 0.9826 & 0.9329 & 0.9330 \\
 \hline    
$\pi/4$ & 0.9662 & 0.9319 & 0.9003 & 0.9003 \\
        &        & 0.9664 & 0.9337 & 0.9338 \\
        &        & 0.9693 & 0.9365 & 0.9365 \\
        &        & 0.9779 & 0.9448 & 0.9448 \\
        &        & 0.9826 & 0.9494 & 0.9494 \\
 \hline    
$\pi/6$ & 0.9829 & 0.9319 & 0.9159 & 0.9159 \\
        &        & 0.9664 & 0.9499 & 0.9499 \\
        &        & 0.9693 & 0.9527 & 0.9527 \\
        &        & 0.9779 & 0.9612 & 0.9612 \\
        &        & 0.9826 & 0.9658 & 0.9658 \\\hline 
\end{tabular}
\end{center}
\label{table9}
\end{table}

In table \ref{table9}, we present the results obtained for a
system of masses 0.1-10$M_{\odot}$ and different choices of $\iota$.
For each case we calculate the fitting factor between the signal
and the family of templates; then, using different parameter
values ${\gras \lambda}_1$ and $ {\gras \lambda}_2$, 
we calculate the  match in the
restricted PN plane $M (u({\gras \lambda}_1), u({\gras \lambda}_2))$,
and, for the same
parameters, the match between the signal and the template
$M (h({\gras \lambda}_1), u({\gras \lambda}_2))$.

What we observe is that the \lq \lq real match" can be approximated
as
$M (h({\gras \lambda}_1), u({\gras \lambda}_2))\approx FF\times
M (u({\gras \lambda}_1), u({\gras \lambda}_2))$.

We turn now attention to the issue of estimating the source parameters.
It is important to notice that
for a waveform computed taking into account PN corrections to the
amplitude, 
two additional parameters are involved; our choice corresponds
to $F_{\times}/F_+$ and $\cos\iota $.
We adopt here the standard 
variance-covariance matrix analysis \cite{Cu94,Po95}, 
although
it can underestimate the statistical errors in the limit of low SNR
\cite{iucaa,nv98}; such well-known problem is beyond the purposes of this work
and for sake of simplicity we will remain into the usual frame of
the computation of the Fisher information matrix. 
 
\begin{table}
\caption{Measurement errors at SNR=10.}
\vspace{0.4cm}
\begin{center}
\begin{tabular}{|l| ccccccc|}
\hline
 & $\Delta A/A$ & $\Delta t_c$ & $\Delta \phi_c$ & $\Delta F_{\times}/F_+$ &
 $\Delta \cos\iota $ &  $\Delta  \M /\M$ & $\Delta \eta/\eta$\\ \hline
 Restricted PN & 0.1000 &  3.3281$\times 10^{-3}$ & 8.0357 & - &  - &
 	 8.4256$\times 10^{-3}$ &0.1107 \\ 
0.5 PN - 2 PN & 0.1062& 2.1179$\times 10^{-3}$& 4.7571& 0.9435& 0.4291
& 4.2884$\times 10^{-3}$ & 0.0624 \\ \hline
\end{tabular}
\end{center}
\label{t7b}
\end{table}

 In table \ref{t7b}, we compare the 
errors for SNR=10 between the restricted and the \lq \lq 
non-restricted" PN approximation.
 The results refer to a system of 1.4-20$M_{\odot}$ and 
  $\theta=\pi/6$,
$\phi=\pi/4$, $\psi=\pi/4$, and $\iota=\pi/3$.
The use of more accurate wave-forms leads to
smaller errors (by roughly a factor $\simeq 2$)
in the determination of the masses; however the information
about $\iota$ and $F_{\times}/F_+$ remain very poor.

In table \ref{t9} we provide the correlation coefficients for the
same parameter choice. We notice that the  correlation coefficients are also 
smaller and that the  two new parameters have small correlations 
with the other ones. Notice that the amplitude is now correlated
with the other parameters while it is not the case for the
restricted PN case.

\begin{table}
\caption{Correlation coefficients. (a)	 Restricted 2 PN, (b) 0.5-2PN.}
\vspace{0.4cm}
\begin{center}
\begin{tabular}{|l| rrrrr|}
\hline
 (a) &$\ln A$ & $t_c$ & $\phi_c$ & $\ln \M $ & $\ln \eta$\\ \hline
 $\ln A$ &  1.00000 &   0.00000 &   0.00000   &  0.00000  &   0.00000 \\
 $t_c$ &  0.00000  &  1.00000 &   0.99202 &    0.92966  &   0.97681 \\
 $\phi_c$ &  -0.00000  &  0.99202  &  1.00000 &  0.96540  &  0.99576 \\
 $\ln \M $ &   0.00000  &  0.92966   & 0.96540 &  1.00000   & 0.98411 \\
 $\ln \eta$ & -0.00000  &  0.97681  &  0.99576  & 0.98411 &   1.00000 
\\ \hline

\end{tabular}
\end{center}
\vspace{0.4cm}
\begin{center}
\begin{tabular}{|l| rrrrrrr|}
\hline
 (b) &$\ln A$ & $t_c$ & $\phi_c$ & $F_{\times}/F_+$ & $\cos\iota $& 
$\ln \M $ & $\ln \eta$\\ \hline
   $\ln A$ & 1.00000 & 0.04913  & 0.04649 & -0.00075 & 0.33173 & 0.03815 
    &  0.04394\\
   $t_c$ &   0.04913  &  1.00000  &  0.96733 &  -0.13841 &  -0.00999  &  
   0.87264   & 0.95706\\
   $\phi_c$ &    0.04649  &  0.96733 &   1.00000  &  0.05476  & -0.01925 
    &  0.92878  &  0.98488\\
 $F_{\times}/F+$ &   -0.00075 &  -0.13841  &  0.05476   & 1.00000  & 
 -0.00209  & -0.06104  & -0.06759\\
  $\cos\iota $&     0.33173 &  -0.00999  & -0.01925  & -0.00209  &  
  1.00000  & -0.02954  & -0.02390\\
  $\ln \M $ &  0.03815  &  0.87264   & 0.92878 &  -0.06104  & -0.02954 &  
   1.00000   & 0.96912\\ 
 $\ln \eta$ &    0.04394   & 0.95706  &  0.98488  & -0.06759  & -0.02390 
   & 0.96912  &  1.00000\\ \hline
\end{tabular}
\end{center}
\label{t9}
\end{table}

\end{document}